# Structural features of PbSe nanoparticles produced within the silicate glasses

V. S. Gurin

*Research Institute for Physical Chemical Problems, Belarusian State University, Minsk, Belarus*

G. E. Rachkovskaya, G. B. Zakharevich

*Belarusian State Technological University, Minsk, Belarus*

S. E. Kichanov, Yu. E. Gorshkova

*Joint Institute of Nuclear Research, Dubna, Russia*

**Abstract**

PbSe-doped silicate glasses were studied through wide-angle X-ray diffraction (WXRD) and small-angle neutron scattering (SANS). PbSe nanoparticles in the glasses were produced due to the secondary heat treatment at different temperatures and their appearance was established by WXRD, SANS, and TEM techniques with the results for particle sizes in accordance to our recent optical studies of this material. The simulation of SANS measurements revealed new structural features of the PbSe nanoparticles. They form fractal aggregates as several overlapping spheres up to ~30 nm in full size.

# 1. Introduction

Quantum-sized semiconductor nanoparticles incorporated into solid matrices are of great interest for design of novel optical materials, non-linear optical units, etc. Optical glasses belong to perspective type of such materials and with lead chalcogenide nanoparticles possess many advanced features for infrared technology, solar energy converters, cutting filters for a range of IR and visible spectra, etc. [1-4]. Lead chalcogenides (PbS and PbSe) are featured by explicit manifestation of quantum size effects through wide size range because they possess the large values of the Bohr exciton radii (18 and 46 nm, respectively) and small effective masses of electron and holes ( _______________ , respectively).

The nanoparticles can be successfully stabilized within a glass matrix of various type (silicate, phosphate, borate). For the size range of 2-10 nm their excitonic absorption peaks enter the near IR range that is of great interest for application in lasers for telecommunication and medicine [5,6]. Meanwhile, the nanoparticles synthesized through soft chemistry methods (colloidal, sol-gel, etc.) may have typically the less size and more tuned particle properties with the optical absorption also in the visible range up to UV. In contrast to colloidal chalcogenides, a material with particles within glass matrices is more stable with respect to environment as well to intense laser irradiation that provides pathways to use glasses as non-linear optical media with proper control of spectral properties via the variation of lead chalcogenide particle state (size, concentration, aggregation, etc.). However, detailed structural studies of nanoparticles in glass are troubled because low particle concentration and background effects due to matrix. Also, any extraction of particles from glass is impossible without their essential modification. Thus, more adequate technique to reveal structural information on nanoparticles is to be non-destructive, e.g. scattering of X-rays and neutrons. Small-angle neutron scattering (SANS) results in versatile structural information on composite nanosystems at different scales as far as neutrons are scattered on various structural inhomogeneities ranging from atomic sizes through tens of nanometers. In the present paper, we combine SANS and wide angle X-ray diffraction (WXRD) for exploration of silicate glasses with PbSe nanoparticles in order to shed a light onto their structural features those influence also optical response of these glasses. The latter are of interest for application as new optical materials for non-linear optical elements, IR-filters, mode-locking units of IR-lasers [ ]. To date, there is no detailed knowledge of structure of such nanocomposite materials since the glass matrix is rather complicated multicomponent system with several oxides, and semiconductor dopants can perturb the glass structure rather significantly as well more structuring of nanoparticles is possible resulting in change of mechanical and optical properties.

Principal optical properties of these glasses are provided by doping with lead chalcogenides. They were presented by us earlier [7] and some similar glasses with PbS and PbSe were described by the other authors [4,8-11]. The doped glasses after the heat treatment at the temperatures near $T_g$ during rather prolonged procedure (20-150 h) reveal the pronounced excitonic absorption as familiar feature of quantum-sized semiconductor nanoparticles in different media. Spectral position of the excitonic absorption is varied in the near IR-range in correspondence with numerical data of the band structure calculations of the nanoparticles in [7]. The PbSe-doped glasses heated at 480°C during 50-60 h correspond to beginning particle growth steps, while the treatment at 525°C exhibits the pronounced peaks in the range of $\lambda > 2$ μm. Thus, within the framework of the present study we consider the couple of samples with different steps of nanoparticles formation and various structural features are expectable to be detected with SANS.

## 2. Experimental: preparation of glasses and measurement techniques

The samples of PbSe glasses were fabricated according to the technique described earlier [7]. It is based on the conventional two-step method used for semiconductor-doped glasses: (1) The melting-cooling cycle of glass-forming oxides ($SiO_2$, $Na_2O$, $ZnO$, $Al_2O_3$, and PbO) with addition of NaF. Elemental Se was used as selenium source. The synthesis temperature was kept $1400 \pm 50$°C during 2 h with the temperature increase rate about 300°C/h. (2) The secondary heat treatment near $T_g$ of the glass (480-530°C). PbSe nanoparticles in this system are nucleated and growing during the second step through a complex mechanism with auxiliary participation of ZnO [12]. Temperature and duration of this step in this protocol is a tool to control final state of the nanoparticles-glass system and its optical features.

XRD patterns for the glasses were recorded using a DRON-7 diffractometer with $CuK_\alpha$ radiation and Ni-filter. TEM study was performed by the method of carbon film replica with extraction. This method provided a view of both particles and glass matrix surface across the sample particular area.

SANS measurements were made using a time-of-flight spectrometer UMO at a IBR-2 reactor (Dubna, Russia). The scattering parameter range covered 0.05-2.5 nm$^{-1}$ allowing to get information on structures at the various scale, 2-100 nm. Raw data of SANS were corrected taking into account thickness of samples and absorption in them and scattering from support and were processed with SASfit software [13].

]Table 1. Composition of the glass under study

| Oxides and other components | weight percent |
|---|---|
| $SiO_2$ | 58.92 |
| $Na_2O$ | 14.56 |
| ZnO | 14.75 |
| $Al_2O_3$ | 4.19 |
| PbO | 3.86 |
| F (NaF) | 2.03 |
| Se | 1.69 |

## 3. Results and discussion
### 3.1. WXRD

The diffraction patterns for the three PbSe-doped glass samples for different conditions of their heat treatment are given in Fig. 1. There are explicit diffraction maxima those can be assigned to appearance of PbSe nanocrystalline phase (JCPDS 06-0354), however, they are of rather low intensity because concentration of PbSe within glasses is low and the nanometer-range particle size provides broadening. This is also a reason that only principal peaks appear, and another possible peaks at $2\theta > 50$ deg are not revealed. Also, amorphous glass matrix can mask good diffraction pattern. Nevertheless, the XRD data depicted in Fig. 1 can be considered as sufficient to conclude on formation of PbSe nanocrystals since they are at close positions to the reference data. Another possible nanocrystalline phases in this system those may be supposed with Pb, i.e. metallic Pb, PbO and $PbSe_2$ do not fall in correspondence with observed data. The weaker peaks at about 21 deg for the sample 525/50 are assignable to partial crystallization of $SiO_2$ (cristobalite phase, JCPDS 39-1425). The assignment of these patterns for the cubic (rocksalt) PbSe phase is given by the reflections (111), (200) for both samples and else (222) for the sample 525/50. The reference positions are 25.16; 29.14 and 41.68 deg, respectively. A comparison of the XRD data for the glasses under study with similar research published in [8-11] indicates good agreement and confirm conclusion on formation of nanocrystalline PbSe with conventional rocksalt lattice.

However, in our case we can see the slight shift in position of the (111) peak while the other ones are better coincident with the reference of bulk PbSe. This observation can argue on small distortion of the nanoparticles structure from the perfect cubic lattice assigned to this space group (Fm-3m, No. 225). While, under ambient conditions only this phase of PbSe is stable, the transitions to the other structures is well known under high pressures and elevated temperatures [14]. The restricted number of diffraction peaks available in the present experiment, their low intensity and broadness do not allow to proceed accurate refinement procedures to decode the distorted lattice of the nanoparticles, however, one can assume the effect of glass matrix in which the particles tightly fixed in the result of melting-cooling cycles. E.g. in the PbS-TiO$_2$ system where the concentration of PbS nanoparticles is higher to resolve the structure through electronographic measurements [ ], the distortion of PbS cubic lattice has been established and simulated by DFT calculations. In our case, the particle-matrix interaction may be even stronger due to the high temperature of the glass preparation and multi-oxide composition of glass. Such interaction is expected to result in strong stabilization of nanoparticles that is of interest for non-linear optical application of these materials.

The full range XRD patterns in the appropriate scale include also broad haloes in the 2θ interval of 20-30 deg that is typical for silicate glasses of different composition and assigned to features of silicate amorphous structure (due to slightly varied Si-O distances) rather than a crystalline doping phase.

Approximate particle sizes from these XRD patterns can be evaluated through the conventional Scherrer method estimating the peaks width, and the average diameter of particles is about 10 nm. A low intensity of the peaks in these patterns does not allow to conclude certainly on difference of the particle size for two samples using this approximate method, but little variations may not be excluded (not more than 1-2 nm). Meanwhile, the difference of XRD diffraction patterns for the two samples consists in more clear appearance of the peaks (~29 and ~41 deg) for the sample 525/50. That can be due to more number of nucleated particles under the higher temperature rather than their larger size. This factor can influence the structural organization of particles contributed into results of SANS measurements and their interpretation (below).

**3.2. SANS**

The raw data of SANS measurements (Fig. 2) display the scattering curves for the two samples under study. These scattering curves appear to be rather different in the full experimental range of scattering parameter Q that should be taken into account in elaboration of models for each glass. A search of possible models corresponding to the scattering data for these glasses issued from the knowledge of their structure as near-spherical nanoparticles distributed in low concentration (low amount of Pb and Se in the precursors, Table 1) within the amorphous glass matrix.

In order to find adequate models applied throughout the full range of these SANS measurements we consider separately three intervals of Q: $Q<0.02$, $0.02 \leq Q \leq 0.1$ and $0.1<Q<0.3$. The fitting of a single model in the full range of Q was not successful. Within the framework of capabilities of SASfit software [13] we incorporated into the simulation both fractal aggregates of overlapping spheres and isolated spheres. For fractal aggregates the scattering function the following expression including the form-factor (spherical particles in this model) and structural factor (general form now for some aggregates):

$$I(Q) = 4\pi \int_0^\infty g(R) R^2 \, \frac{\sin(QR)}{QR} dR,$$

The function g(R) describes the aggregates through the diameter, $2\xi$, that characterizes the dimension of space (d=3 in this model), and D is the value of fractal dimension. g(R) function is defined for this type of aggregates as

$g(R) = R^{D-d}((1+R/4\xi)(1-R/2\xi)^2)$       for $R<2\xi$
$g(R) = 0$      for $R \geq 2\xi$.

The fitting of experimental data for the two samples under study to this model is presented in Fig. 2a,c.

In the next range of scattering parameters, $0.02 \leq Q \leq 0.1$, we use the Guinier approximation

$I(Q) = (4\pi^2 N^2/3)\exp(-Q^2 R_g^2/3),$

where N is particle concentration and $R_g$ is the gyration radius which is related to particle radius for the spherical case as $R=(5/3)^{1/2}R_g$. The result of this fitting is given in Fig. 2b,d (the part of smaller Q, QR<1) and decoded as presence of spherical particles of certain radius (R).

The fitting at the larger Q in these plots corresponds to the generalized Porod law [ ] in which we account both surface fractals ($D_s$) and mass fractals (D) in the scattering intensity:

$$I(Q) = c_0 + c_4 Q^{D_s-2D}$$

where $c_4$ is the fitting constants and the term $c_0$ accounts the background. In our case we obtain by the simulation the exponent $D_s-2D$ close to the value -4, i.e. the Porod law if to consider only one value for the dimension parameter. Taking into account possibility of different $D_s$ and D, the surface fractal dimension $D_s$ appears to be different for the two samples since the mass fractal dimension, D, covered two values (Table 2) for these two samples. For the first sample, 480/56, $D_s$=1.84 that means the almost smooth surface for scattering objects, while, in the case of the sample 525/50 the surface features are more complicated, $D_s$=1.06. These results can evidence that the temperature-stimulated modification of PbSe nanoparticles affects the particle-glass interface. The effect may be associated with pre-melting the glass matrix at the $T>T_g$ with slight growth of the nanoparticles size providing changes at the glass-particle interface.

Table 2. Parameters of the models obtained through fitting procedures of SANS data for two samples of PbSe-doped glasses

| Sample | $\xi$, nm | D | $D_s$ | $R_g$, nm | R, nm |
|---|---|---|---|---|---|
| 480/56 | 27.6 | 2.92 | 1.84 | 6.13 | 7.9 |
| 525/50 | 24.5 | 2.53 | 1.06 | 8.11 | 10.5 |

Table 2 summarizes the data obtained through the above analysis of SANS data for two glass samples. They indicate that structure of glasses may be presented as combination of single particles ($R_g$~6-8 nm) and fractal aggregates of different dimension. The size of aggregates is of the order of tens of nanometers and comprised of several particles of the size range R~8-10 nm. The value D is close to 3 (dense aggregate) for the 480/56 sample while the 525/50 sample reveals formation of less dense aggregates and possesses more complicated surface. That can means slight growth of PbSe nanoparticles under the temperature rise (480/56 → 525/50) accompanied by disintegration of aggregates.

### 3.3. TEM

A typical micrograph of nanoparticles within glass (480/56 sample) is presented in Fig. 4. We observe both single particles of the size range of 5-20 nm and little amount of aggregates those are comprised of two or several particles. Locations of each particle or aggregate demonstrate specific areas around them those may be associated with changes in glass matrix during the cooling process as far as the melting temperature of PbSe is higher than $T_g$ for the glass. This picture is typical for different semiconductor-doped glasses with the glass matrix of similar composition (alkali-silicate) doped with CdTe, CdSe, PbS, $CuInSe_2$, etc.), since usually a semiconductor material and glass matrices have different thermal expansion features. Inhomogeneous density of glass can contribute slightly into the above neutron scattering patterns, however, it is hardly decodable from the experimental data together with PbSe nanoparticles since the scattering cross section by the latter is much higher than any contribution due to matrix density variations. Thus, the size range and morphology of these particles due to PbSe-doping in the results of TEM, in general, are in consistence with the above conclusions from SANS data simulation.

### 4. Conclusions

In this work, we studied the PbSe-doped silicate glasses in which PbSe nanoparticles have been formed due to the secondary heat treatment at different temperatures. Appearance of the nanoparticles localized within glass matrix was established by WXRD, SANS, and TEM techniques with results in accordance to previous optical studies of this material. The size range of the particles from the measurements with different techniques is evaluated in the range of 5-10 nm. The simulation results of SANS reveal new structural features of the PbSe nanoparticles: they form fractal aggregates as several overlapping spheres up to ~30 nm in full size. Thus, the glasses under study are not simple 'particle-in-a-matrix' material. An effect of the structuring upon optics and the nature of aggregation are the subject of our future research.

**Figures captions**

Fig. 1. Diffraction patterns for two samples of the PbSe-doped glasses labeled by temperature and time of the second heat treatment.

Fig. 2. Raw data of SANS measurements for two samples of the PbSe-doped glasses labeled by temperature and time of the second heat treatment.

Fig. 3. Fitting of the SANS data for two samples of PbSe-doped glasses through different ranges of the scattering parameter, Q: (a-b) 480/56 and (c-d) 525/50 (temp, $^{o}$C/time of treatment, h)

Fig. 4. Transmission electron microscopy image of the PbSe-doped glass, the 480/56 sample.

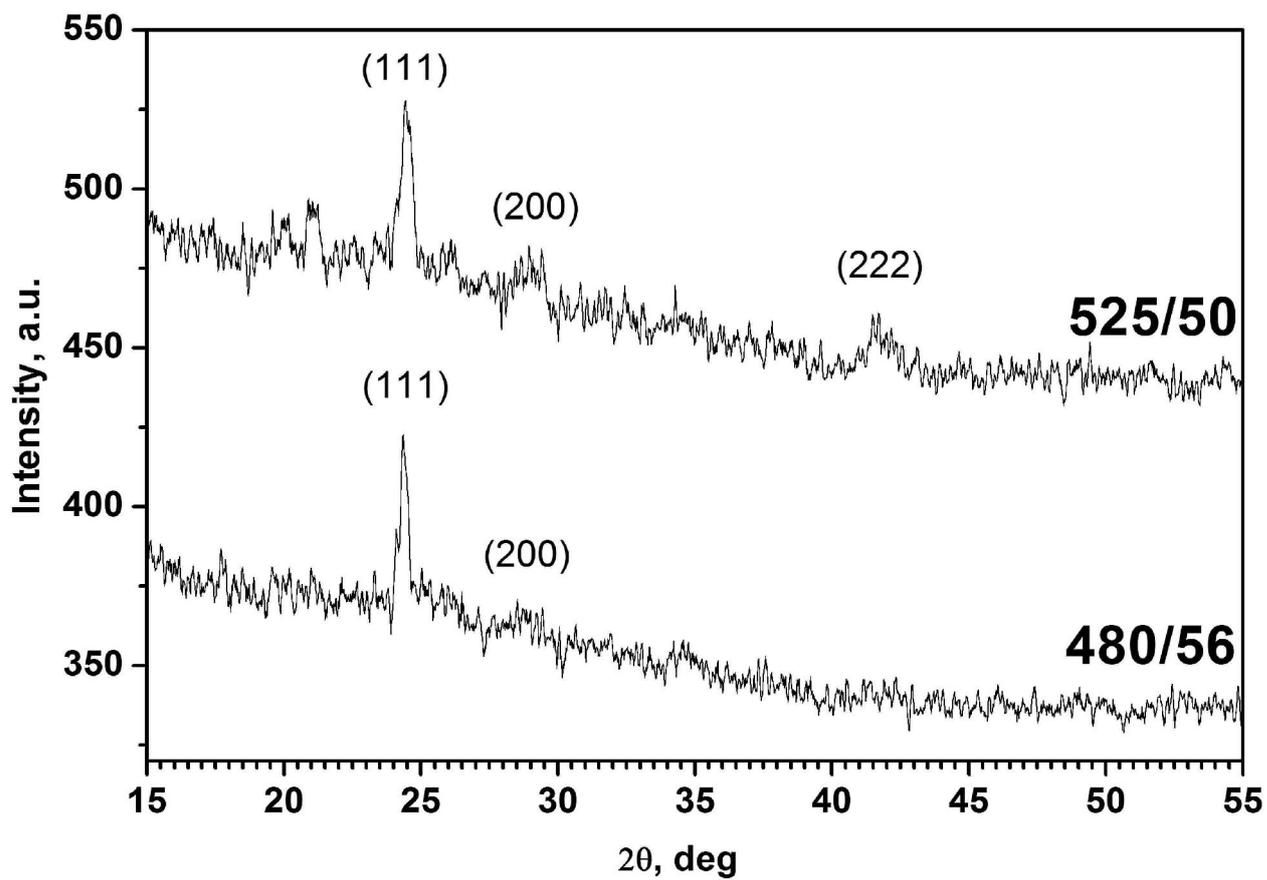

Fig 1

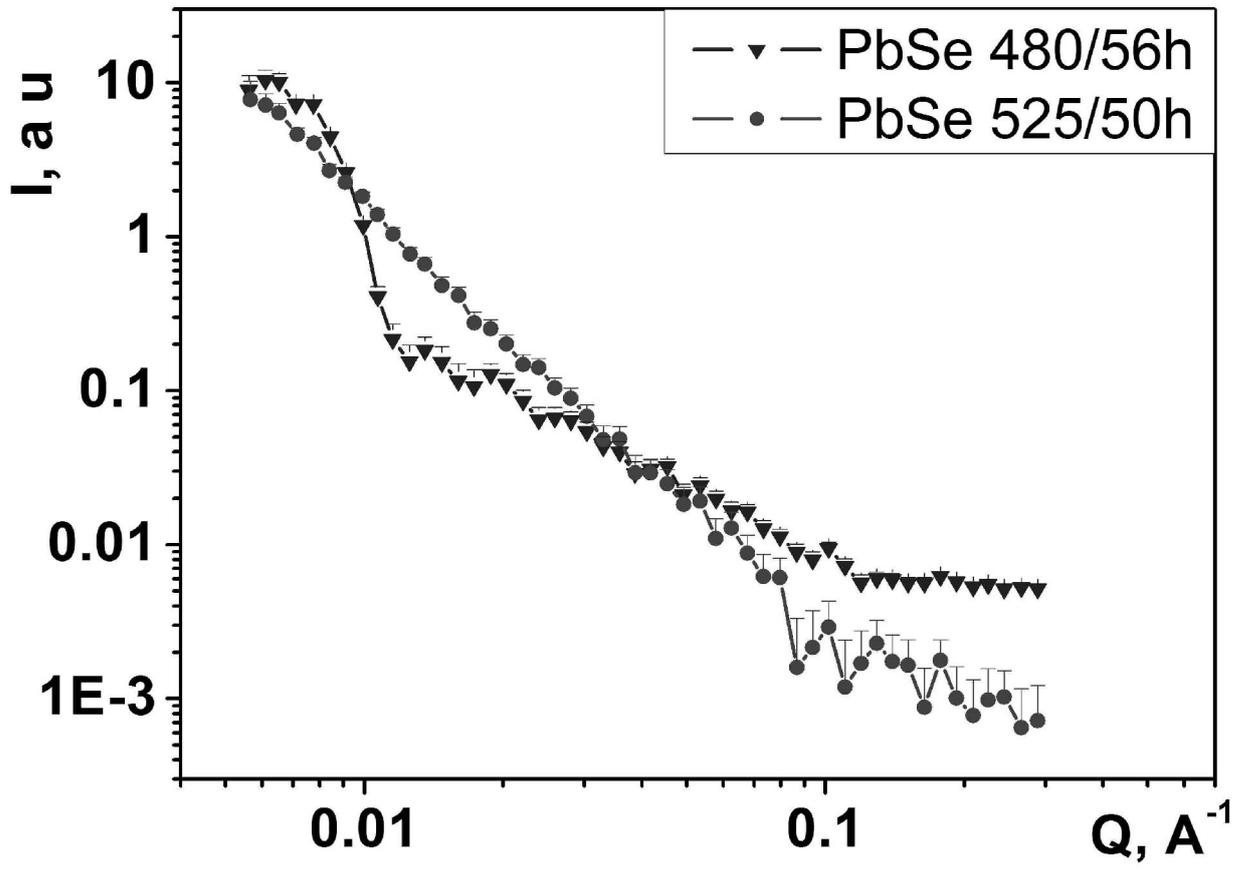

Fig 2

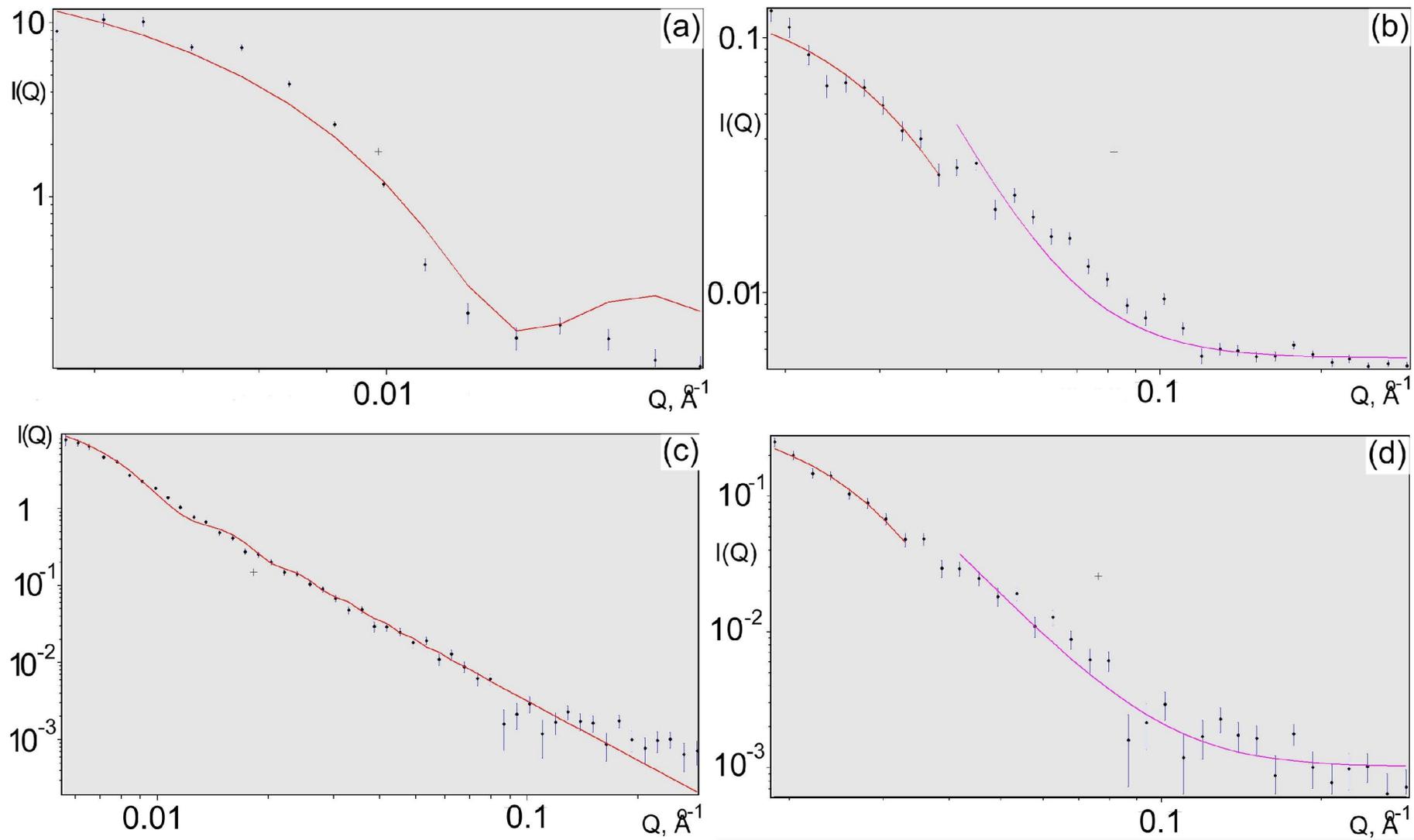

Fig 3

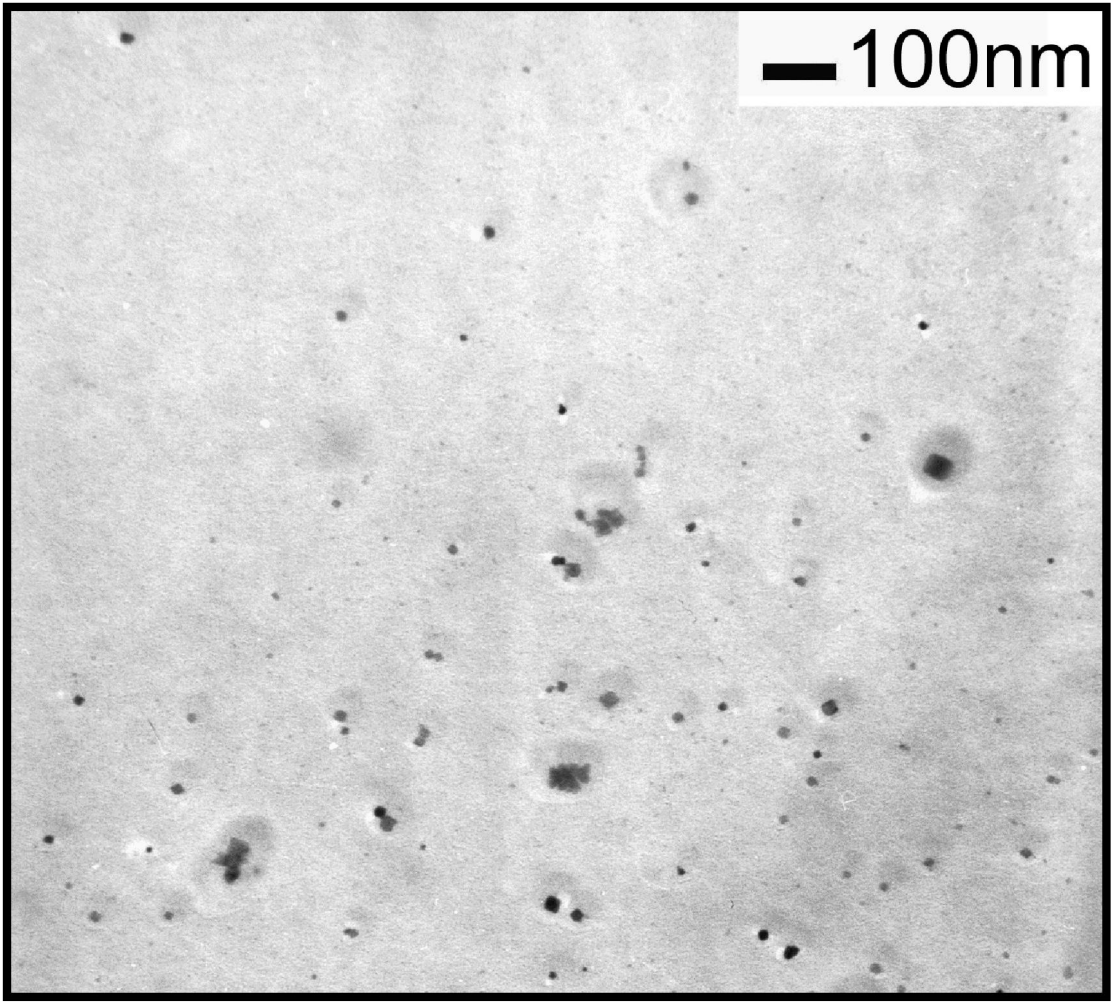

Fig 4